\documentclass[]{aa}
\usepackage{psfig}
\begin{document}

\thesaurus{20(08.15.1, 08.16.4)}

\title{Multiperiodicity in semiregular variables}
\subtitle{II. Systematic amplitude variations}

\author{L.L. Kiss\inst{1} \and K. Szatm\'ary\inst{1} \and Gy. Szab\'o\inst{1}
\and J.A. Mattei\inst{2}}

\institute{Department of Experimental Physics and Astronomical Observatory,
JATE University,
Szeged, D\'om t\'er 9., H-6720 Hungary \and
American Association of Variable Star Observers (AAVSO), 25 Birch
Street, Cambridge MA02138-1205 USA}

\titlerunning{Systematic amplitude variations in semiregular variables}
\authorrunning{Kiss et al.}
\offprints{l.kiss@physx.u-szeged.hu}
\date{}

\maketitle
 
\begin{abstract}

We present a detailed lightcurve analysis for a sample of
bright semiregular variables based on long-term (70--90 years)
visual magnitude
estimates carried out by amateur astronomers.
Fundamental changes of the physical state (amplitude and/or
frequency modulations, mode change and switching) are studied
with the conventional Fourier- and wavelet analysis.

The light curve of the carbon Mira Y~Per showing
a gradual amplitude decrease has been re-analysed
after collecting and adding current data to earlier ones.
The time scales of the sudden change and convection are compared and their
similar order of magnitude is interpreted to be a possible hint for
strong coupling between pulsation and convection. The periods
of the biperiodic low-amplitude light curve and their ratios
suggest a pulsation in the first and third overtone modes.
An alternative explanation of the observed behaviour could be
a period halving due to the presence of weak chaos.

Beside two examples of repetitive mode changes (AF~Cyg and W~Cyg)
we report three stars with significant amplitude modulations
(RY~Leo, RX~UMa and RY~UMa).
A simple geometric model of a rotationally induced amplitude modulation
in RY~UMa is outlined assuming low-order nonradial oscillation,
while the observed behaviour of RX~UMa and RY~Leo
is explained as a beating of
two closely separated modes of pulsation. This phenomenon is detected
unambiguously in V~CVn, too. The period ratios found in these stars
(1.03--1.10) suggest either high-order overtone or radial+non-radial
oscillation.

 
\end{abstract}

\section{Introduction}

Semiregular stars (SRVs) of type SRa and SRb are
pulsating red variables on the
asymptotic giant branch (AGB) of the Hertzsprung-Russell diagram.
According to the definition in the General Catalogue of
Variable Stars (GCVS), their visual amplitude does not
exceed 2.5 mag and the light variation is not as strictly
cyclic as in Mira stars.
The attribute ``semiregular'' refers to the complexity of their
light curves, which has been usually interpreted as being caused
by cycle-to-cycle
variations in a relatively simple pulsating environment.
There has been an increasing amount of empirical
evidence that multiperiodicity occurs is some stars (Mattei et al. 1998,
Kiss et al. 1999 -- Paper I), which suggests the existence of many
simultaneously
excited modes of pulsation. Lebzelter (1999) and Lebzelter et al. (2000)
presented radial velocity measurements and their comparison with
simulteneous light curves for several bright semiregular
variables, which confirmed that main periodicities are most likely
due to pulsation and partly due to large-scale motions in the extended
atmospheres of red giants.

From the observational point of view, the main parameters characterizing the
pulsation are as follows: periods and amplitudes, the number of
excited modes, their nature (radial
or non-radial) and orders (in case of overtone pulsation).
Fundamental changes of these
parameters (e.g. mode-switching, repetitive turning on and off of different
modes, long-term amplitude or frequency modulations) may indicate incomplete
understanding of the physical processes governing the pulsation and stellar
evolution. Mode-changes were reported in several stars
(Cadmus et al. 1991, Percy \& Desjardins 1996, Bedding et al. 1998,
Kiss et al. 1999), however, no firm conclusion was drawn about the
responsible physical mechanism.
Since these red giants have extended
convective envelopes around the stellar cores, earlier linear stability
surveys (e.g. Fox \& Wood 1982, Ostlie \& Cox 1986), ignoring or
oversimplifying the coupling between convection and pulsation, may suffer
from serious incompleteness. Xiong et al. (1998) performed
linear survey treating the dynamic and thermodynamic
couplings by using a statistical theory of nonlocal and
time-dependent convection and outlined the ``Mira'' pulsational
instability region (see also Gautschy 1999).
Their calculations support the possible
existence of high-order (up to 3rd and 4th) overtone modes, which
has been suggested by empirical studies, e.g. Percy \& Parkes (1998),
Kiss et al. (1999).

\begin{table*}
\begin{center}
\caption{The list of programme stars. Variability types are taken
from the GCVS, while periods are those of obtained in Paper I.}
\begin{tabular}{|l|l|l|l|l|}
\hline
GCVS    &    IRAS       &   Type    &  Dataset(s)                   & Periods \\
        &               &           &  [JD+2400000]                 & [days] \\
\hline
RS Aur  &   --          &   SRa     &  28485 -- 50890 (AFOEV, VSOLJ) & 173, 168\\
V CVn   &   13172+4547  &   SRa     &  37600 -- 51208 (AAVSO)        & 194, 186\\
        &               &           &  23808 -- 50907 (AFOEV, VSOLJ, HAA/VSS) &         \\
W Cyg   &   21341+4508  &   SRa     &  17404 -- 50902 (AFOEV, VSOLJ, HAA/VSS) & 240, 130\\
AF Cyg  &   19287+4602  &   SRb     &  22456 -- 50902 (AFOEV, VSOLJ, HAA/VSS) & 921, 163, 93\\
RY Leo  &   10015+1413  &   SRb     &  28598 -- 50933 (AFOEV, VSOLJ, HAA/VSS) & 160, 145\\
Y Per   &   03242+4400  &   M       &  37602 -- 51210 (AAVSO)        & 245, 127\\
        &               &           &  23057 -- 51547 (AFOEV, VSOLJ, HAA/VSS) &         \\
RX UMa  &   09100+6728  &   SRb     &  17689 -- 50808 (AFOEV, VSOLJ, HAA/VSS) & 201, 189, 98\\
RY UMa  &   12180+6135  &   SRB     &  37600 -- 51208 (AAVSO)        & 305, 287\\
\hline
\end{tabular}
\end{center}
\end{table*}

The first part of this series addressed the general properties
of semiregular variables of types SRa and SRb based on a set
of semiregulars containing almost 100 stars. A few special cases
were reported to illustrate such phenomena as long-term amplitude
modulations, amplitude decrease and mode switching. In the meantime,
new and newly computerized archive data were added to those analysed
previously; therefore, some of the special cases in Paper I are worth
studying in more details. The main aim of this paper is to discuss
the observed changes of the pulsational properties through well-observed
light curves in terms of multimode pulsation, possible non-radial
modes, coupling between oscillation and/or rotation and convection.
Also, we want to extend the list of stars with such phenomena and
to point out that the discussed behaviours are much more common
in red semiregulars than was thought earlier.
The paper is organised as follows. Sect.~2 deals with the observations
and basic data handling, Sect.~3 contains the results on the individual
variables, while a brief summary is given in Sect.~4.

\section{Observations}

The presented results are based on long-term visual observations for eight stars
carried out by amateur astronomers. The data were extracted from the
same international databases as in Paper I (Association Francaise
des Observateurs d'Etoiles Variables -- AFOEV, Variable Star Observers'
League of Japan -- VSOLJ, American Association of Variable Star Observers --
AAVSO, Hungarian Astronomical Association/Variable Star Section -- HAA/VSS).
In the meantime newly computerized historic and very recent data, collected
by the AAVSO\footnote{\tt http://www.aavso.org} and
VSNET\footnote{\tt http://www.kusastro.kyoto-u.ac.jp/vsnet},
have been added to the earlier observations. The data were analysed with
the conventional Fourier and wavelet analysis after calculating
5-day or 10-day bins of visual light curves depending on the length of the
most characteristic period (see details and references in Paper I).
The list of programme stars and details on the datasets are
presented in Table\ 1.

\section{Results on the individual variables}

\subsection{Changing the type of variability: Y~Persei}

\begin{figure}
\begin{center}
\leavevmode
\psfig{figure=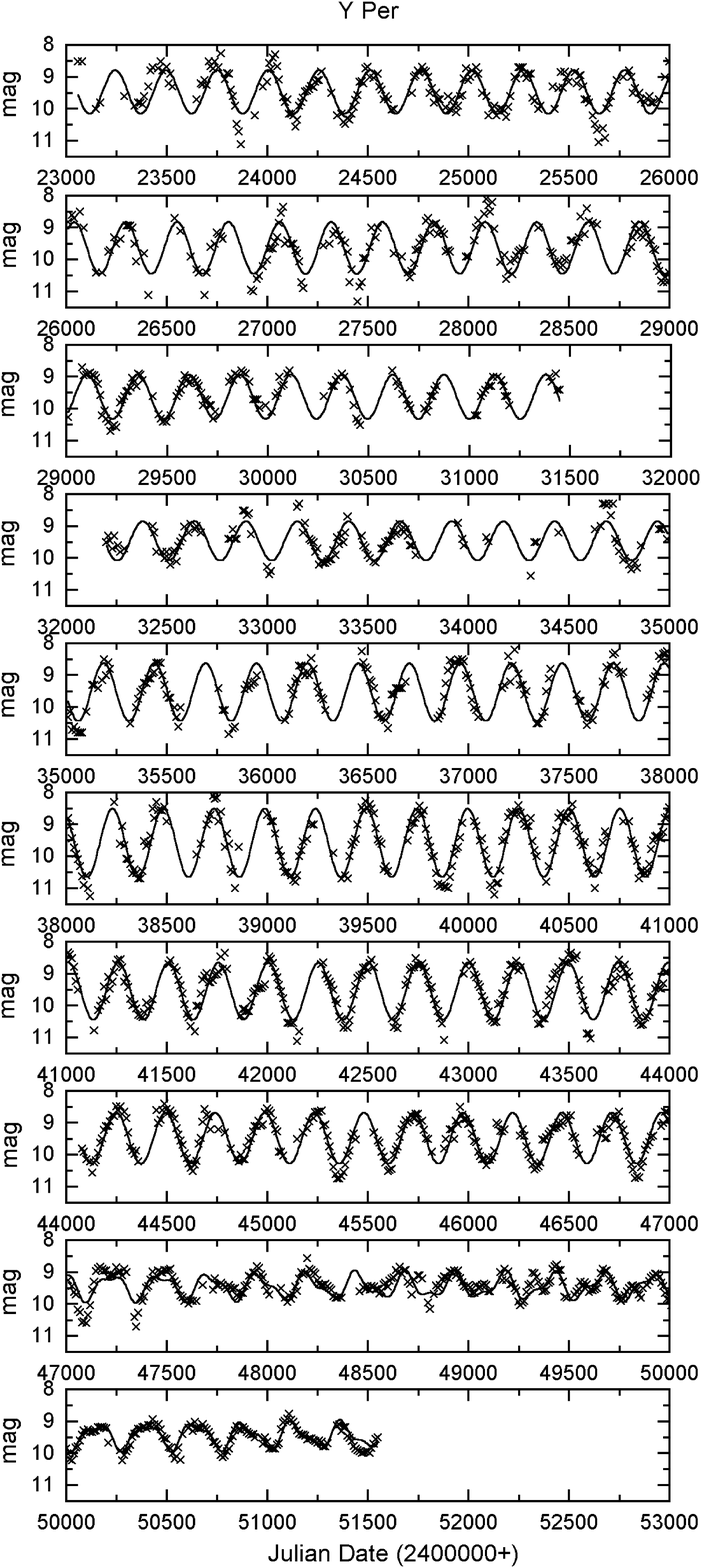,width=\linewidth}
\caption{The lightcurve of Y~Per in 8+1 segments.
The last one is plotted in two separate plots for clarity.}
\end{center}
\label{f01}
\end{figure}

Y~Per is a well-known carbon Mira star, though its period
(about 250 days)
is the shortest one among them (Groenewegen et al. 1998). There was
no indication of peculiarity until 1987, when its amplitude dropped
significantly. Furthermore, it fits exactly the PL relation of
galactic carbon LPVs (Bergeat et al. 1998), thus this star
has been considered as a typical member of its type.

However, as has been pointed out in Paper I,
the appearance of the visual
light curve changed dramatically around JD 2447000 (1987).
The earlier Mira-type variations disappeared and were replaced
by a semiregular and low-amplitude brightness change.
In order to trace the time-dependent variations and
quantify the sudden change, we performed a
detailed lightcurve analysis utilizing subsets
of the whole dataset.

The finally merged and averaged light curve covers 28500 days (more than
110 cycles). There are smaller gaps in the fist half of the data, while
the second half is completely continuous. We divided the data into
eight 3000 day-long segments (each containing about 12 cycles)
and one 4500 day-long segment. This enabled
an accurate period determination in every segment avoiding the
possible period smearing due to its long-term variation.
The period was calculated with the conventional Fourier-analysis and
was checked independently with a non-linear regression analysis.

Data in the first eight segments can be described very well with only
one harmonic, but the last one with a two-component harmonic sum.
None of the residuals shows significant periodic signals.
The fitted curves are plotted in Fig.\ 1, while the resulting parameters
are presented in Table\ 2. Although the formal standard errors
are quite small, the real uncertainties are a little larger, most
probably due to the intrinsic variations of Y~Per. Therefore,
we adopted a period uncertainty of $\pm$1 day and an amplitude error
of $\pm$0.05 mag.

\begin{table}
\begin{center}
\caption{Frequencies, the corresponding periods
and amplitudes in nine subsets.}
\begin{tabular}{llll}
\hline
No.     &    $f (10^{-3}c/d)$   &  $P (day)$  & $A (mag)$\\
        &    (s.e.)             &             & (s.e.)   \\
\hline
1       &    3.950 (0.080)      &  253.2      & 0.67 (0.09)\\
2       &    3.915 (0.009)      &  255.4      & 0.81 (0.03)\\
3       &    3.954 (0.007)      &  252.9      & 0.69 (0.03)\\
4       &    3.930 (0.012)      &  254.5      & 0.63 (0.04)\\
5       &    3.953 (0.006)      &  253.0      & 0.91 (0.03)\\
6       &    3.959 (0.004)      &  252.6      & 1.07 (0.03)\\
7       &    4.023 (0.005)      &  248.6      & 0.89 (0.02)\\
8       &    4.058 (0.005)      &  246.4      & 0.80 (0.02)\\
9       &    4.095 (0.005)      &  244.2      & 0.36 (0.02)\\
        &    7.864 (0.012)      &  127.2      & 0.16 (0.02)\\
\hline
\end{tabular}
\end{center}
\end{table}

The amplitude and period changes seem to be well correlated
as shown in Fig.\ 2., which is a well-known property of
non-linear oscillators. A further interesting point is that
the secondary period occuring in the last segment is
exactly the half of the earlier dominant period within the
error bars (the mean period in the first six segments is 253.8 days).
Although it can be pure numeric coincidence, we will shortly
discuss the possible relevance of this period halving.

As has been mentioned in Paper I, the abruptness of this
amplitude change is quite surprising. The characteristic
time scale of the amplitude decrease in other stars
(V~Boo - Szatm\'ary et al. 1996, R~Dor - Bedding et al. 1998,
RU~Cyg - Paper I) varies from few hundred to few thousand days,
typically tens of pulsation cycles. In Y~Per the change happened
in 400 days, between JD 2447200 and 2447600, or less than 2 cycles.
 Adopting a pulsational approach, it can be interpreted as a fast
appearance of a new mode beside a slightly changing dominant mode.
Recent models by Xiong et al. (1998) suggest a first and third
overtone combination (observed periods: 245 and 127 days;
theoretical prediction: P$_1$=231 days, P$_3$=130 days).
According to Xiong et al. (1998), the coupling between
convection and pulsation depends critically on the ratio
of the timescale of convective motion and that of pulsation.
The effect is stronger for overtone oscillation, as
the turbulent viscosity becomes the main damping mechanism
of the high overtones (it converts the kinetic energy
of ordered pulsation into random kinetic energy).
There have been a considerable number of theoretical
and observational efforts in order to quantify
this coupling and to detect observables related to it.
Anand \& Michalitsanos (1976) have already formulated a simple
nonlinear model assuming that the convective envelope of
M giants is composed of giant convection cells
comparable in size to the stellar radius. They showed that the
coupling can produce asymmetric fluctuations of the
entire star. Such large perturbations may cause also
fast changes in the pulsational properties. The earlier
theoretical models are supported by modern high-resolution
spectroscopic observations. Nadeau \& Maillard (1988)
observed velocity gradients of a few km~s$^{-1}$ for lines
of different excitation potentials, which was interpreted
as being caused by convective motions. Most recently,
Lebzelter (1999) presented
such radial velocity measurements of semiregular variables,
which implied the possibility of large convective cells with
radial motions close to a few km/s.
A simple estimate of the convective time scale can be found from
the ratio of the thickness of the convective envelope and the
observed order of magnitude of convective velocity.
For 100 R$_\odot$ and 1 km~s$^{-1}$ one get a result
of 800 days being close to the duration of sudden
changes in SRVs.
Unfortunately, present instability
surveys cannot calculate the amplitude of pulsation and,
consequently, its long-term variation can be interpreted
only speculatively.

\begin{figure}
\begin{center}
\leavevmode
\psfig{figure=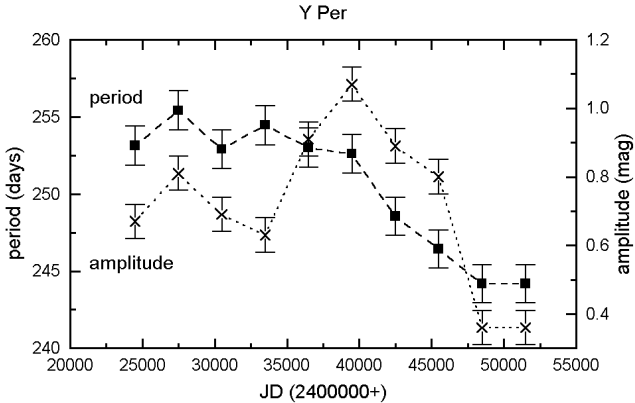,width=\linewidth}
\caption{The variations of the dominant period and amplitude
in the 3000 day-long segments}
\end{center}
\label{f02}
\end{figure}

We note that the first and third overtone model is very similar
to that derived for R~Dor by Bedding et al. (1998). They concluded
in this instance that the rapid changes can be described
very well with chaotic effects discussed by Icke et al. (1992).
Furthermore, the period halvings noted above might be another
hint for the presence of weak chaos. It can be interpreted as
an inverse process of the period doubling bifurcation which
may happen just beyond the onset of chaos
(Kov\'acs \& Buchler 1988). It has been shown in a number of papers
(e.g., Kov\'acs \& Buchler 1988, Saitou \& Takeuti 1989,
Moskalik \& Buchler 1990),
that different stellar pulsation models (W~Vir, RV~Tau, yellow
semiregulars) show period doubling bifurcation leading from a regular
to a quite irregular variation. Although the behaviour of Y~Per
does not fit exactly the proposed way of transition from
regular to irregular state through cascades of period doubling
bifurcations (and simultaneous noisy period halvings),
this kind of explanation cannot be excluded.

Finally, the conventional classification of long period
variables (LPVs) into Mira and SRa, SRb, and SRc type
semiregular variables, as given in the GCVS, is not
based on physical parameters and does not adequately cover
even the behavior of some of the brightest
and most studied LPVs.
This has been improved by Kerschbaum \& Hron (1992)
by involving the `blue' and `red' subgroups of semiregulars.
The SRa stars seem to form a mixture of intrinsic Mira and SRb variables.
For Mira stars no similar definition using various stellar parameters exists.
Unfortunately, the fact that Y~Per does not fit neither of main types
of LPVs (Mira and semiregular), does not uncover the underlying
physical processes being responsible for its peculiarity.

\subsection{Amplitude modulation I.: beating in RX~Ursae Majoris and
RY~Leonis}

In Paper I we presented eight stars with closely separated frequencies.
A few of them are triply periodic stars, in which mode changes occur
quite frequently (see Sect. 3.4) and the close frequency components
do not exist simultaneously. However, we found two variables with
stable biperiodic variation, where various subsets show the same
frequency content (V~CVn and RY~Leo). In those cases the amplitudes of
components are quite different,
therefore, no clear beating occurs, as in RX~UMa. We plotted a typical
subset of the light curve of RX~UMa in Fig.\ 3 where a three-component
fit is also shown (f$_0$=0.005000 c/d, A$_0$=0.37 mag, f$_1$=0.005288 c/d,
A$_1$=0.26 mag, f$_2$=0.010236 c/d, A$_2$=0.16 mag). The frequency
spectrum of the whole dataset is plotted in Fig.\ 4.

\begin{figure}
\begin{center}
\leavevmode
\psfig{figure=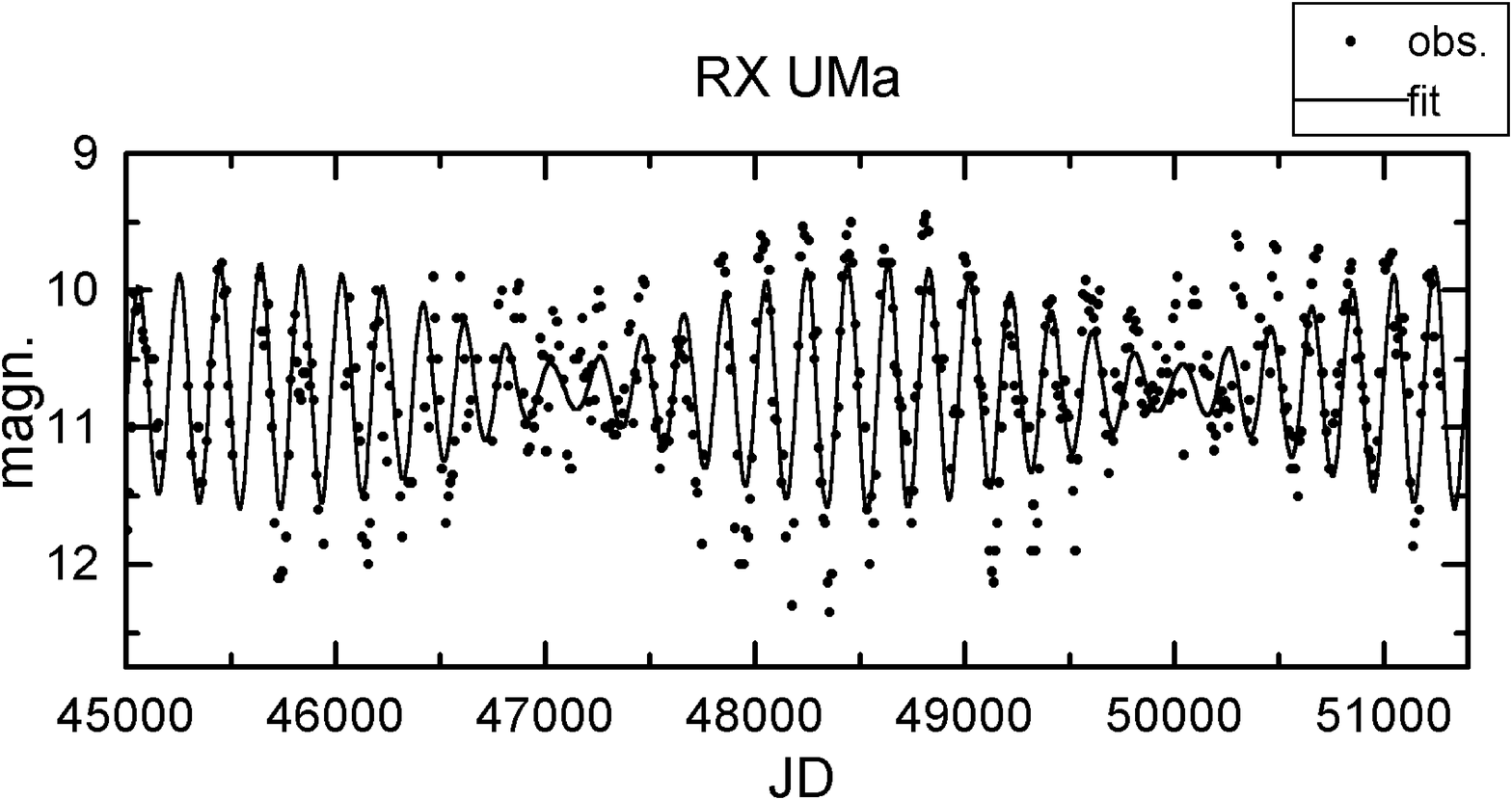,width=\linewidth}
\caption{The observed and fitted light curves of RX~UMa.}
\end{center}
\label{f03}
\end{figure}

\begin{figure}
\begin{center}
\leavevmode
\psfig{figure=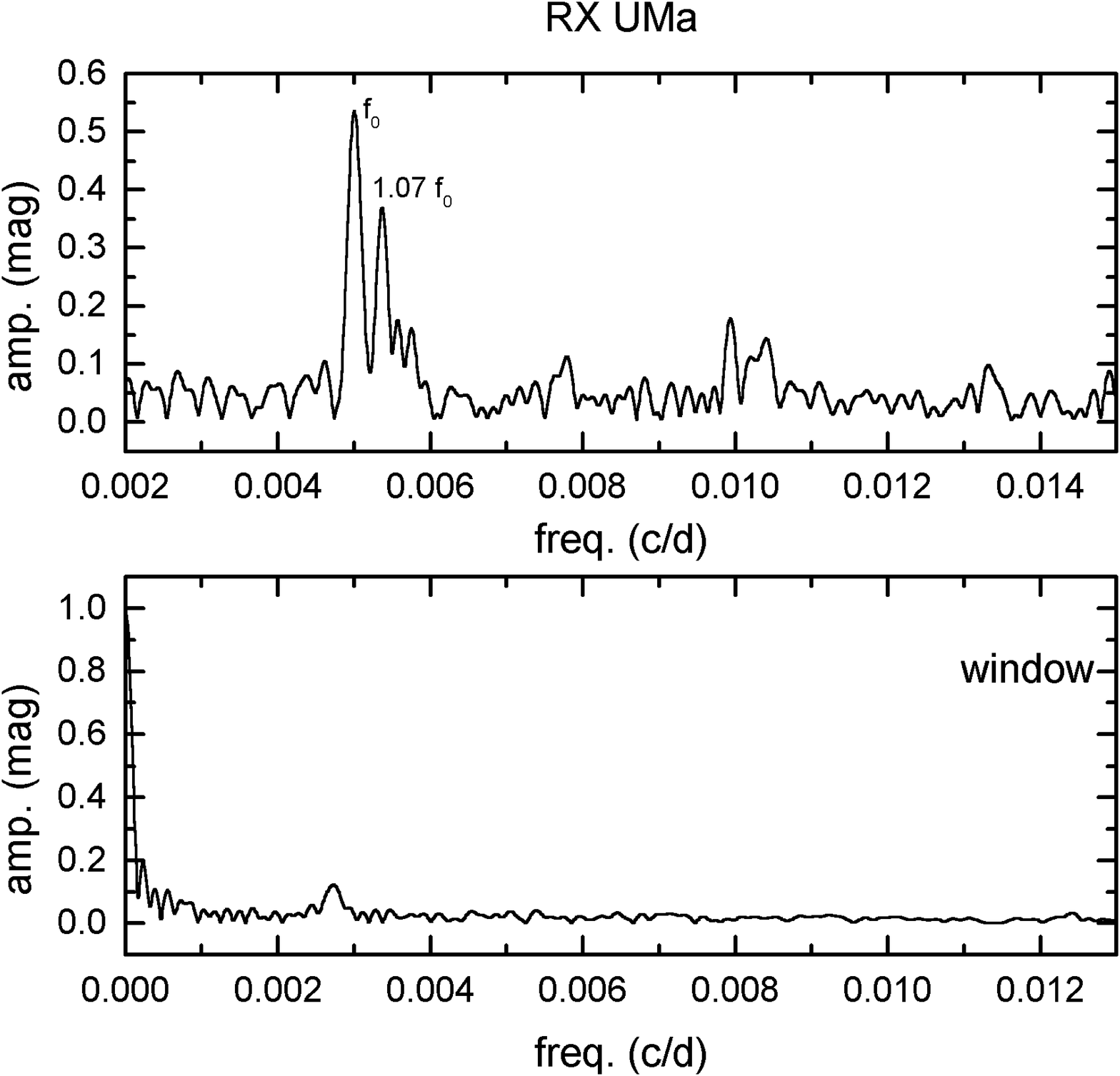,width=\linewidth}
\caption{The frequency spectrum ({\it top})
and window spectrum ({\it bottom}).}
\end{center}
\label{f04}
\end{figure}

If these frequencies have pulsational origin, then their ratios
(f$_1$/f$_0$=1.06, f$_2$/f$_1$=1.93) may give information about the
mode of pulsation. Period ratios near 1.9 are very common in
semiregular variables (see Paper I and Mattei et al. 1998) which
was identified in case of R~Dor to correspond to the first and
third overtones (Bedding et al. 1998). Period ratios close to 1
may suggest high (3-5th) overtones, but theoretical calculations
suggest strong damping in this mode domain (Xiong et al. 1998).
Mantegazza (1988) found similar close frequency doublet in the
red semiregular Z~Sge and using models by Fox \& Wood (1982)
speculated about the possibility of second and third overtones.
It is also possible, that one of the frequencies correspond to
a non-radial mode (e.g. Loeser et al. 1986), but not much is known 
about the non-radial
pulsation of red giant stars. In the other two stars mentioned earlier
there are also two closely-spaced periods, which are demonstrated by
the DFT spectra in Figs.\ 5-6. Note, that in V~CVn even the
cross-production terms (f$_0$$\pm$f$_1$) are present unambiguously.

\begin{figure}
\begin{center}
\leavevmode
\psfig{figure=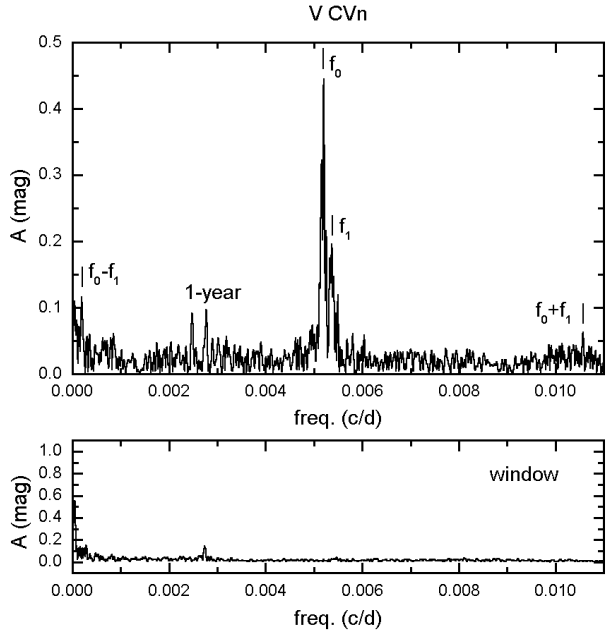,width=\linewidth}
\caption{The frequency spectrum of V~CVn.}
\end{center}
\label{f05}
\end{figure}

\begin{figure}
\begin{center}
\leavevmode
\psfig{figure=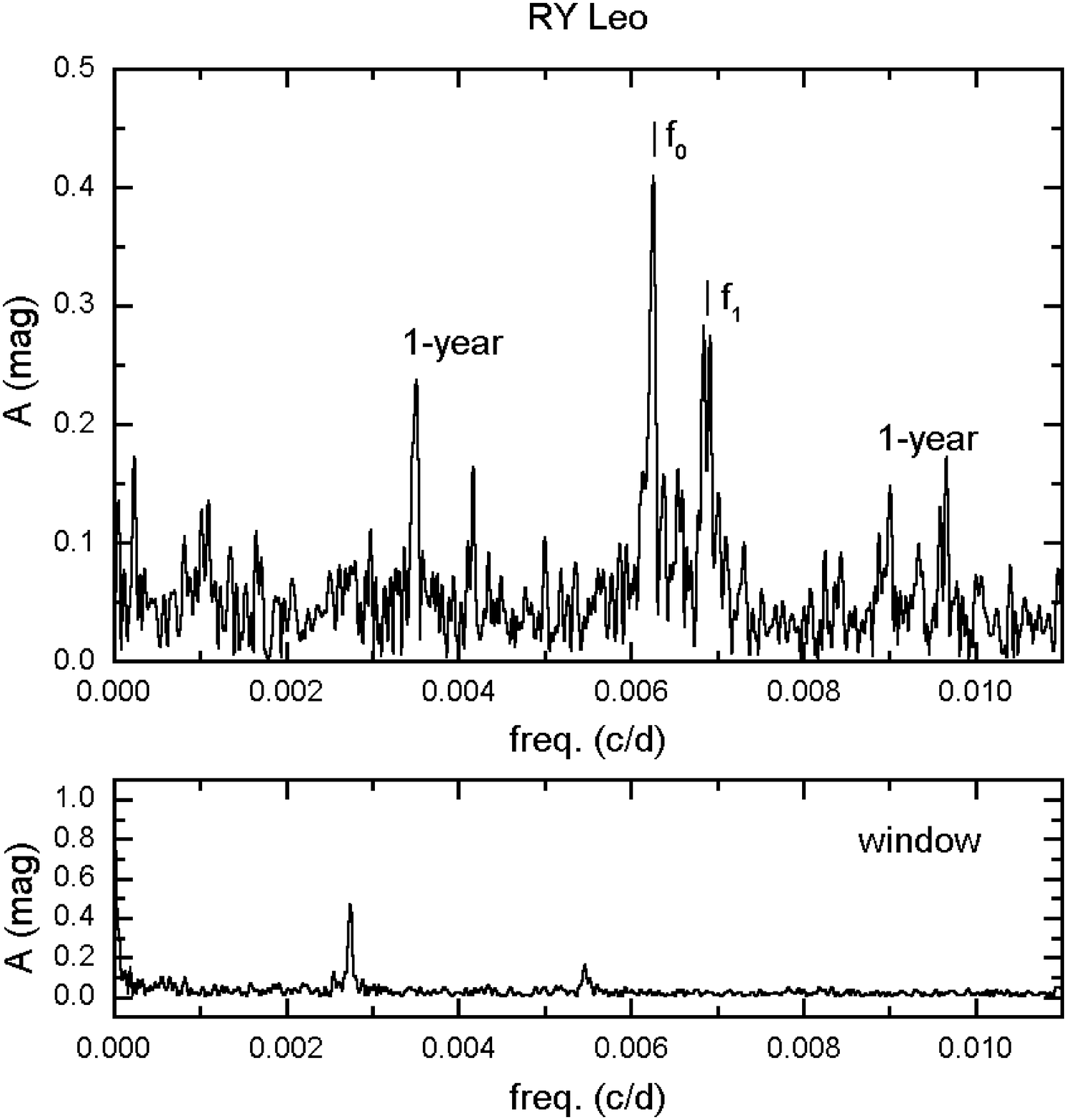,width=\linewidth}
\caption{The frequency spectrum of RY~Leo.}
\end{center}
\label{f06}
\end{figure}

\subsection{Amplitude modulation II.: pulsation + rotation in RY~Ursae~
Majoris?}

\begin{figure*}
\begin{center}
\leavevmode
\psfig{figure=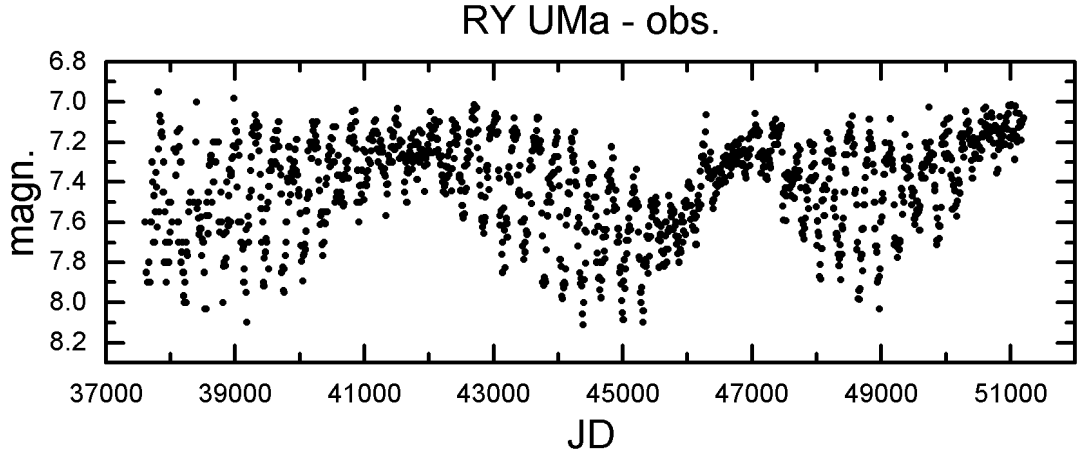,width=16cm}
\caption{The observed light curve (10-day means) of RY~UMa.}
\end{center}
\label{f07}
\end{figure*}

The amplitude modulation discussed in the previous subsection
can be simply described by the
beating of two close frequencies, since the mean brightness did not
change with time in those stars. However, RY~UMa (and partly RS~Aur)
shows amplitude variations which highly resembles those
observed in RR~Lyrae variables with Blazhko effect,
where the minimum brightness
changes much more significantly,
than the maximum brightness. Therefore, a
significant mean brightness variation can be observed.

This interesting light curve variation has been highlighted in
Paper I, where a 9800-day segment was analysed. In the
meantime, historical AAVSO observations were added, extending the
light curve to a whole length of 17000 days. Unfortunately, the
early light curve during the first 3000 days has a less dense coverage,
thus the most homogeneous data cover about 14000 days. The corresponding
averaged
light curve is shown in Fig.\ 7. In our sample of 93 stars studied
in Paper I this behaviour is quite rare, only RS~Aur seems to have
similar light curve phenomenon (Fig.\ 8).

We have tried to explain the observed amplitude modulation with rotational
effects. Earlier theoretical studies have generally neglected the
stellar rotation, since typical rotational periods of red
giants, usually obtained theoretically, are about 4000--10000 days,
much longer than the characteristic times of pulsation. However,
RY~UMa shows such complex light variation and frequency spectrum
(see later), that
a possible explanation could be the rotation-pulsation connection.
The amplitude variations turned out to be highly repetitive, which is
shown in Fig.\ 9, where the light curve has been folded with two
periods (P$_{\rm pul}$=306 days, P$_{\rm mod}$=4900 days$\approx 16\cdot$P$_{\rm pul}$).

\begin{figure}
\begin{center}
\leavevmode
\psfig{figure=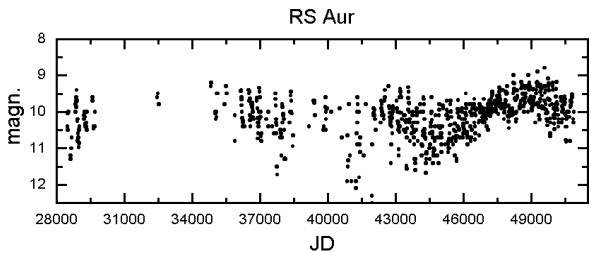,width=\linewidth}
\caption{The observed light curve (10-day means) of RS~Aur.}
\end{center}
\label{f08}
\end{figure}

\begin{figure}
\begin{center}
\leavevmode
\psfig{figure=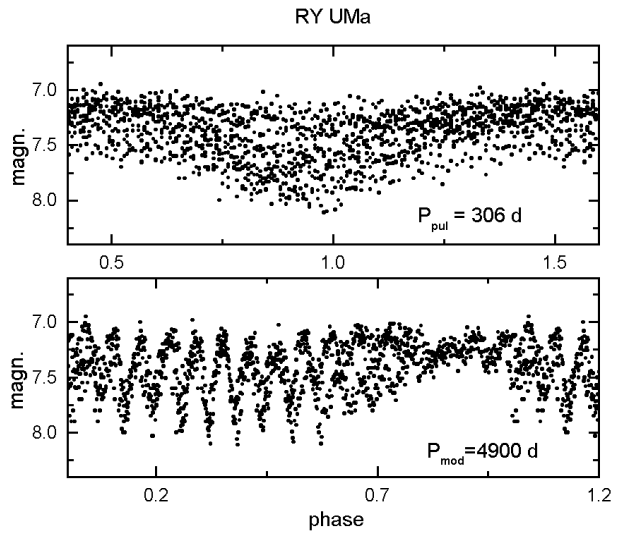,width=\linewidth}
\caption{The light curve of RY~UMa folded with the periods of
pulsation ({\it top}) and modulation ({\it bottom}).}
\end{center}
\label{f09}
\end{figure}

We have tried to build a simple model, which involves a rotationally
modulated non-radial oscillation. Our very approximate model consists of:
{\it i)} a distorted stellar shape caused by a low-order non-radial
oscillation; {\it ii)} stellar rotation with a period of 9800 days
(i.e. twice of the period of modulation); {\it iii)} simple
limb darkening (u=0.6). We have considered a triaxial ellipsoid having
a short axis of unity, while the other axes change sinusoidally
in time between 1.0--2.0 and 1.75--2.0 with the period of pulsation (306 days).
The whole ovoid rotates and the intensity is integrated over the surface
elements assuming normal limb darkening with coefficient u=0.6. The light
curve is calculated as the logarithm of the change of the surface
facing the observer.

All aspects of this approach can be, of course, easily challenged.
The introduced distorsion in our model is much larger than that of
predicted by traditional description of non-radial oscillations.
However, there have been a number of high-resolution observations
(imaging and interferometry) of nearby Mira variables showing
substantial asymmetric structures (Tuthill
et al. 1999, Lopez et al. 1997, Karovska et al. 1997), which imply
the incompleteness of spherical assumptions. Furthermore,
fully three dimensional and turbulent dynamic numerical
simulations of red giant stars (Jacobs et al. 1998) also
suggest occurence of bipolar atmospheric motions and distorted
stellar (photospheric) shape. Thus, we conclude, that it {\it may be}
possible that the assumed maximum oblateness for RY~UMa is not
completely unlikely.

The second aspect is stellar rotation. Asida \& Tuchmann (1995)
explored theoretically the asymmetric mass-loss from rotating red
giant variables and presented a scenario for an anisotropic
mass ejection from AGB variables caused by rotational effects.
Further support of rotationally induced variations in red giants
was given by Barnbaum et al. (1995), who suggested a possible
connection between rapid rotation (P$_{\rm rot} \approx$ 530 days)
and pulsation in the carbon star V~Hya (type SRa). In our
case there is no need for assuming unusual rotation, since
the required rotational period (9800 days) is in the range of
what we expect for such extended and evolved objects as red giants.

\begin{figure}
\begin{center}
\leavevmode
\psfig{figure=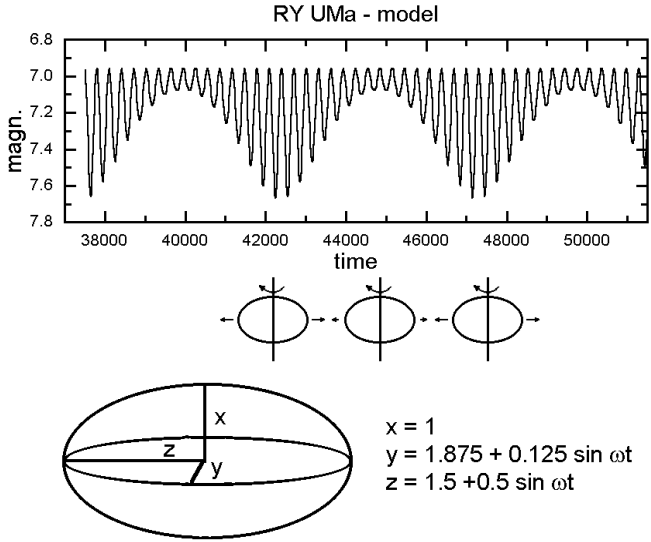,width=\linewidth}
\caption{The calculated model light curve and selected phases with
different geometric aspects.}
\end{center}
\label{f10}
\end{figure}

The weak point in our model is the neglect of temperature
variations along the pulsation and the assumption of
constant (and solar) value for the limb-darkening coefficient. The
latter is less significant, because even completely neglecting
the limb darkening (i.e. using a uniform disk)
does not change the calculated light curve
significantly. This is especially fortunate keeping in mind
dynamical model analyses by Beach et al. (1988), performed
in order to determine limb darkening/brightening
function for Mira atmospheres, which illustrated
that the usual limb darkening correction of uniform disk model
in lunar occultation
measurements can be even in the wrong direction in certain pulsational phases.
The temperature variations could not be
taken into account, since we have no information neither about its
range nor about its phase dependence. The strong non-radial
assumption would imply weaker temperature effects, but without
any kind of phase dependent temperature measurements, we cannot
draw a firm conclusion.
Nevertheless, our main purpose is rather to get a qualitative
``fit'' of the observed behaviour than to quantitatively model
a visual light curve.
A graphical representation of the model and the resulting light curve
is given in Fig.\ 10.

\begin{figure}
\begin{center}
\leavevmode
\psfig{figure=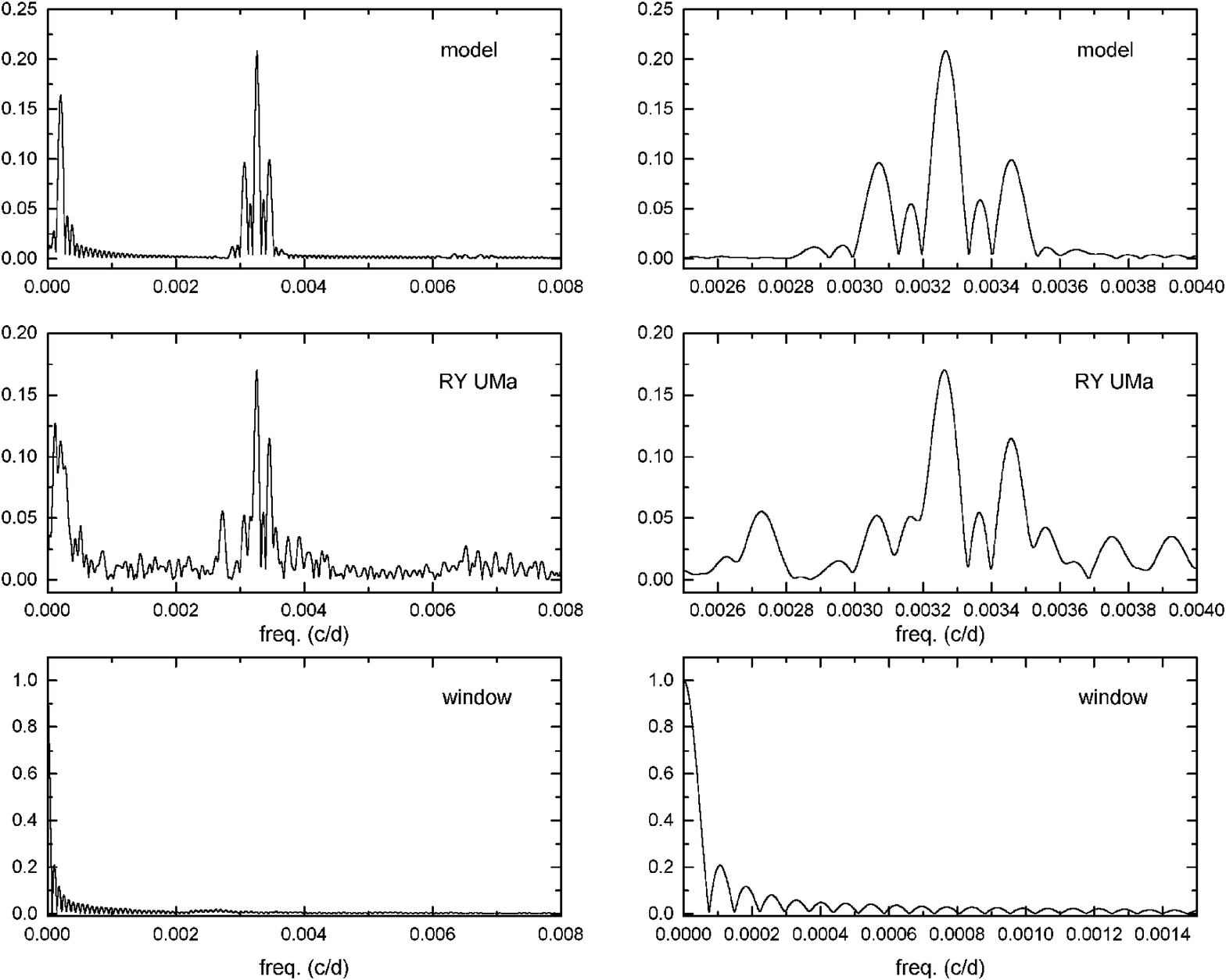,width=\linewidth}
\caption{Discrete Fourier-transforms of the modelled ({\it top}) and
observed ({\it middle}) light curves. Bottom panels show
the window functions. There are apparent splittings
around the corresponding peaks in the frequency spectra, which
are (model) and might be (observations) due to the stellar
rotation.}
\end{center}
\label{f11}
\end{figure}

The resulting model light curve was compared with the observed one
through their frequency spectra. This comparison can be seen in Fig.\ 11,
where the corresponding Fourier spectra are shown.
Beside the overall similarity the most striking feature
is the frequency splitting of the main component at f$_0$=0.003268 c/d.
Such a splitting has been well-known behaviour in pulsating white
dwarfs and roAp stars (e.g. Shibahashi \& Saio 1985, Buchler et al. 1995,
Kurtz et al. 1996, Baldry et al. 1998),
while recently it has been detected in several
RRab stars showing Blazhko effect (e.g. Kov\'acs 1995, Nagy 1998,
Chadid et al. 1999)
and in a number of MACHO RRc variables (Alcock et al. 2000).
One of the commonly accepted views is the oblique pulsator model, where
a non-radial oscillation is coupled with the rotation, as
the rotational axis does not coincide with the symmetry axis of
pulsation (Shibahashi 1999). In the case of RY~UMa, a frequency triplet
is present with $\Delta f \approx$0.0002 c/d, which correspond to
the frequency of amplitude modulation. Also, the triplet has
an asymmetric amplitude distribution with
respect to the central frequency peak. Similar asymmetry can be attributed
to intrinsic nonlinear mechanisms (Buchler et al. 1995), that seems
to be quite likely in a highly non-spherical environment.
We note the presence of a peak at 0.00273 c/d corresponding
exactly to one year. This is little surprising, because the
light curve has no seasonal gaps (it is circumpolar from Europe, Japan
and North America), therefore a such alias peak is not expected.
A possible reason for this could be the seasonal increase of the
scatter, since despite its circumpolarity it is more difficult to observe
near the horizon.

We give further illustration of the amplitude variations by
the wavelet transform shown in Fig.\ 12. This time-frequency analysing
method is quite successful in quantifying the time-dependent
variations in astronomical time series (see, e.g., Szatm\'ary et al.
1994, Szatm\'ary et al. 1996, Foster 1996 and Paper I).
The overall pattern is quite systematic suggesting the
regularity of the underlying process(es). The amplitude
of the main ridge has been extracted along the time axis
in order to plot its variations in Fig.\ 13. Our conclusion is
that despite the local irregularities of the visual
light curve (see the larger scatter in the bottom panel
of Fig.\ 9 between $\phi$=0.6--0.8), the amplitude modulation
implies a regular physical process, e.g. rotation or binarity.
In this paper we adopted to rotation, however, other possibilities
cannot be excluded using the presently available observations.

\begin{figure}
\begin{center}
\leavevmode
\psfig{figure=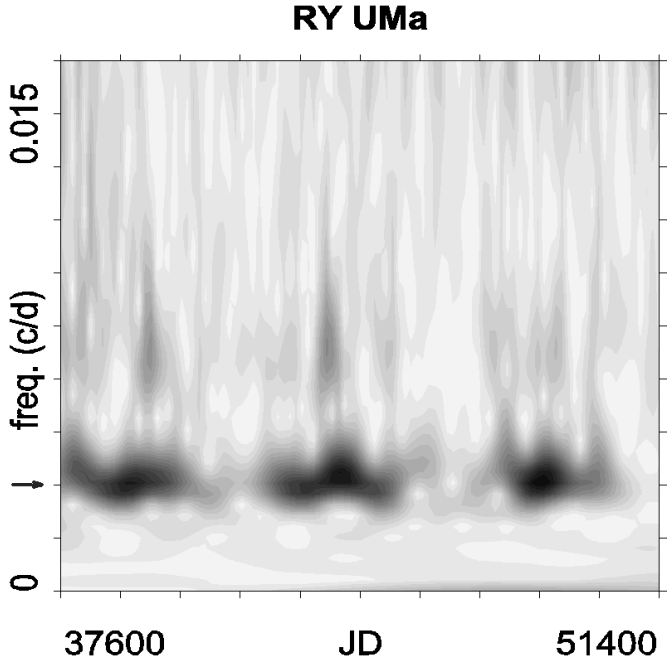,width=\linewidth}
\caption{The wavelet map of RY~UMa. The main ridge of the primary frequency
component (indicated by the small arrow)
was sliced to get the time-dependent amplitude values plotted
in Fig.\ 13.}
\end{center}
\label{f12}
\end{figure}

\begin{figure}
\begin{center}
\leavevmode
\psfig{figure=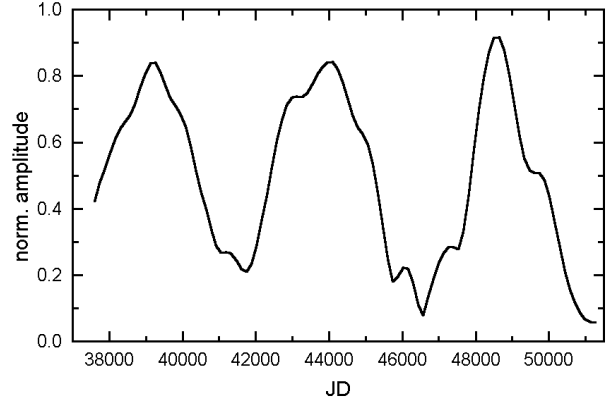,width=\linewidth}
\caption{The normalized amplitude variation of RY~UMa from the wavelet map.}
\end{center}
\label{f13}
\end{figure}

\subsection{Amplitude variations due to repetitive mode changes:
W~Cygni, AF~Cygni}

W~Cyg and AF~Cyg are two of the most popular and well-observed semiregular
stars (e.g. Percy et al. 1993, 1996).
These stars are two illustrative examples of repetitive mode
changes, where some modes turn on and off on a time scale of few
hundreds of days (a few cycles). This has been partly highlighted
in Paper I, where TX~Dra and V~UMi were discussed in terms
of varying dominant modes. In this paper we focus on two other stars,
which further supports our belief that
this phenomenon may be quite frequent in semiregular variables.

\begin{figure}
\begin{center}
\leavevmode
\psfig{figure=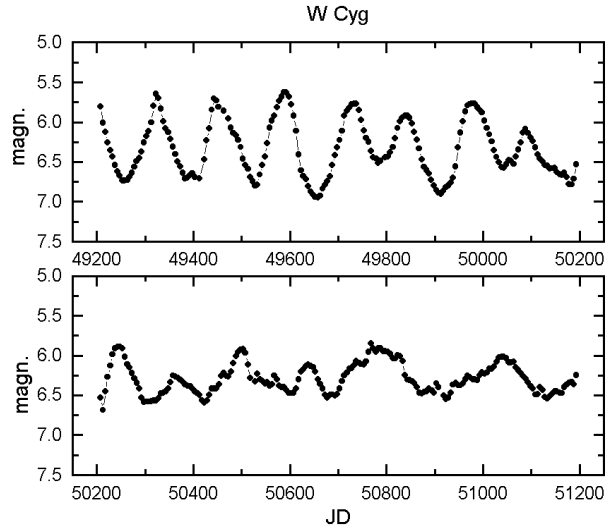,width=\linewidth}
\caption{1000-days long noise-filtered (Gaussian smoothing with an
FWHM of 8 days) subsets of W~Cyg showing different states with
different dominant modes. Note, that solid line only connects the
points and does not represent any kind of fits.}
\end{center}
\label{f14}
\end{figure}

\begin{figure}
\begin{center}
\leavevmode
\psfig{figure=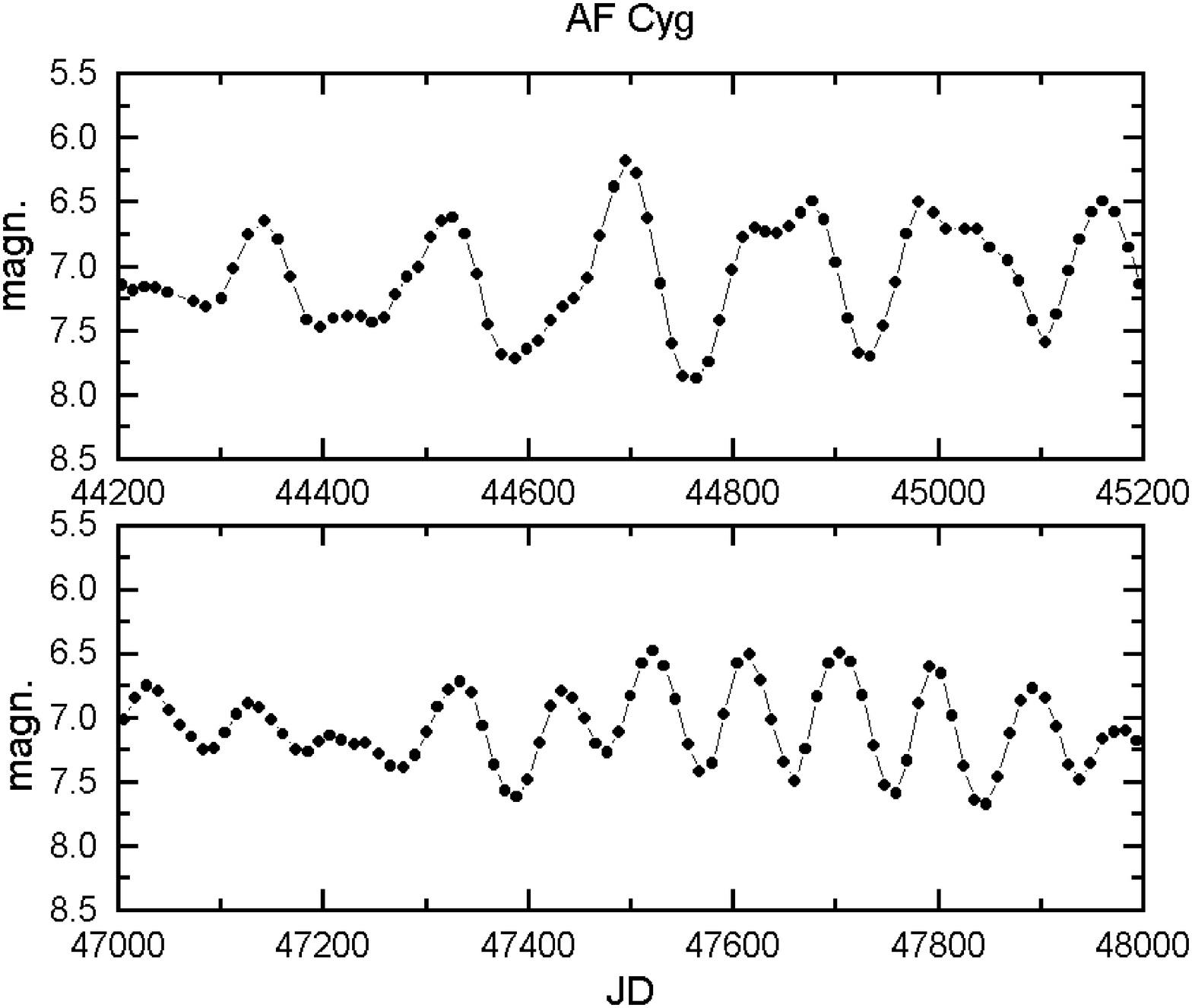,width=\linewidth}
\caption{The same as in Fig.\ 14 for AF~Cyg.}
\end{center}
\label{f15}
\end{figure}

Both stars have continuous light curves covering more than 28000 days,
which corresponds to 300 and 215 cycles for AF~Cyg and W~Cyg, respectively,
considering their shorter periods (93 and 130 days). There are
several occasions when their light curves completely change.
This is illustrated in Figs.\ 14--15, where we plotted two
1000-days long subsets for both stars. The photometric behaviour (cycle
lengths and their amplitudes) changes dramatically from time to time.
The wavelet map of W~Cyg (Fig.\ 16) and the cross sections
of the main ridges (Fig.\ 17) clearly illustrate the amplitude
variations. We obtained similar results with AF~Cyg.

\begin{figure}
\begin{center}
\leavevmode
\psfig{figure=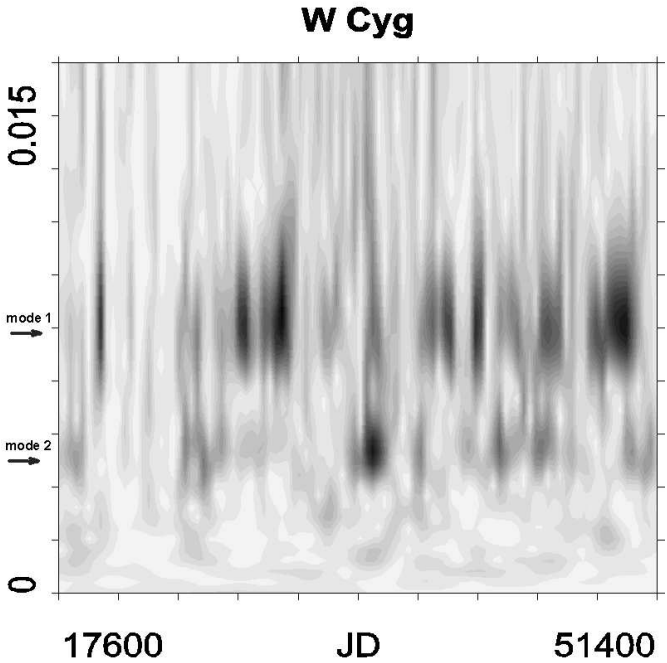,width=\linewidth}
\caption{The wavelet map of W~Cyg.}
\end{center}
\label{f16}
\end{figure}

\begin{figure}
\begin{center}
\leavevmode
\psfig{figure=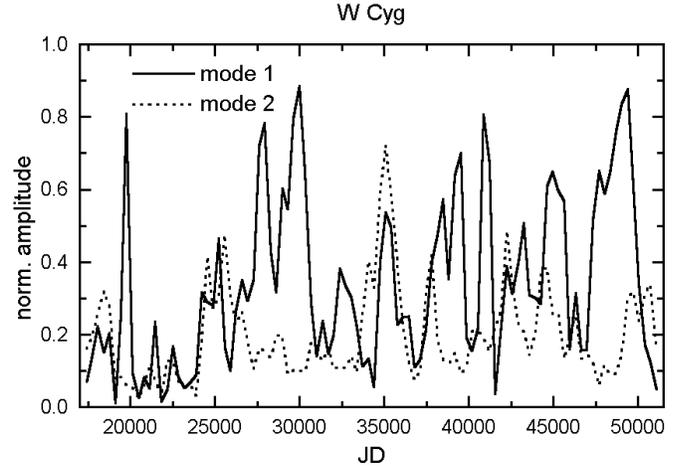,width=\linewidth}
\caption{The normalized amplitude variations of the two main modes.}
\end{center}
\label{f17}
\end{figure}

There are no obvious indications for periodicity in Fig.\ 17.
However, a few important points can be drawn. The first is that the
plotted amplitude changes should be considered real, as the
light curve has no gaps after JD 2421000 (1916). Therefore, the
repetitive amplitude decreases and increases
are not numerical artifacts caused by the inappropriate data distribution
(see Szatm\'ary et al. 1994 for testing the method with simulated
light curves). The amplitude of the dominant mode (``mode 1'', P=130 days)
changes on a time scale of 2000-3000 days. This may imply that
the exciting mechanism of pulsation is intermittent
and the damping is strong. The bimodal state (i.e. simultaneous
high amplitudes of ``mode 1'' and ``mode 2'') is quite rare.
Furthermore, only weak hints are present for simultaneous
or alternating modes suggesting different and independent
excitations for the two modes corresponding to the
first and third overtones in theoretical models of
Xiong et al. (1998) -- P$_1$=231 days, P$_3$=130 days.

As has been mentioned above, AF~Cyg has similar unstable behaviour, only
the periods (and possibly the modes) are different. The two main
periods (163 days and 93 days) would suggest second and
fourth overtones adopting models of Xiong et al. (1998)
(P$_2$=173 or 154 days, P$_4$=103 or 91 days).

Finally, episodic amplitude and period changes were also
reported for three semiregulars (RV~And, S~Aql, U~Boo)
by Cadmus et al. (1991), where
the dominant period was the shorter one during the low-amplitude
epsides. In addition, Mattei \& Foster (2000), who studied
long-term trends of period, amplitude, mean magnitude
and asymmetry in the AAVSO light curves, reported several
stars with such trends. But what we see here is
completely different in W~Cyg and AF~Cyg, where
the shorter period is mainly the dominant one with higher amplitude.

\section{Conclusions and summary}

Our conclusions based on the results presented in
this paper can be summarized as follows.

\noindent 1. We have re-analysed the light curve of the carbon Mira
star Y~Per complemented with the most recent observations. Its variation
has been transformed from a monoperiodic oscillation to a bimodal
one since 1987, which periods suggest first and third overtone pulsation.
The suddennes of this switch is interpreted as a consequence of a strong
coupling between pulsation and convection. We conclude that the
unphysical classification scheme of LPVs (Mira and semiregulars) could be
misleading when speculating on different evolutionary states
of these variables. A simple distinction between monoperiodic
and multiply periodic red variables may be more straightforward in
some cases. Another important consequence of first plus third
overtone assumption is that Y~Per, a typical carbon Mira before 1987,
has pulsated in the first overtone mode. That would imply short period
Miras to pulsate in the first overtone, but no firm conclusion
can be drawn on this issue based on only one star.

\noindent 2. We report two stars (RX~UMa and RY~Leo) with significant
amplitude modulation with no changes of the mean brightness. We
have interpreted this behaviour as beating of two closely
separated periods, most likely high-order radial or radial plus
non-radial oscillation. Similar frequency doublet is reported
for V~CVn, confirming earlier results of Loeser et al. (1986).
Unfortunately, there is no thorough theoretic study
on the non-radial oscillations of red giants so far, thus we could
only speculate on this explanation comparing semiregulars
with other well-documented types of variables with
similar close frequencies (i.e. $\delta$ Scuti and ZZ~Ceti stars).
However, the relatively large fraction of such SRVs in our sample
of 93 stars implies that this phenomenon may be quite common in
these stars.

\noindent 3. We have also found another type of amplitude
modulation resembling RR~Lyrae variables with Blazhko
effects
in RY~UMa and possibly in RS~Aur. Again, since this aspect
has no firm theoretic background yet, we assumed similar
physical mechanism being responsible for similar light variations.
Our qualitative model involving highly distorted stellar shape
describes very well the observed modulation, though the
initial physical assumptions may be considered quite unlikely.
However, there are indications for oblate stellar shapes
from high-resolution observations in several Mira stars
and also there are empirical pieces of evidence for complex
circumstellar structures around semiregular variables,
though mainly observed in the infrared (Bergman et al. 2000,
Kerschbaum \& Oloffson 1999, Knapp et al. 1998).
Further observation with high spatial and spectral resolution
are highly recommended in order to clarify the situation.

\noindent 4. Two bright semiregular stars, W~Cyg and AF~Cyg,
are used to illustrate the repetitive excitation and damping
of main modes of pulsation. The characteristic time scale
is roughly 2000--3000 days. We have shown that only
weak evidence is present for correlated or anticorrelated
amplitude changes of different modes implying the possibility
of independent excitation mechanisms. Theoretical
pulsation models including the coupling between
convection and pulsation were used to identify W~Cyg and
AF~Cyg to be first+third and second+fourth overtone
pulsators, respectively.

\begin{acknowledgements}

We sincerely thank variable star observers of AFOEV, VSOLJ, HAA/VSS and
AAVSO whose dedicated observations over many decades made this
study possible.
This research was supported by Hungarian OTKA Grants \#F022249,
\#T022259, \#T032258 and Szeged Observatory Foundation.
The NASA ADS Abstract Service was used to access data and references.
\end{acknowledgements}

\end{document}